\documentclass[aps,pra,twocolumn,amsfonts,superscriptaddress,showpacs]{revtex4-1}
\usepackage{amsmath,amssymb,mathrsfs}
\usepackage{latexsym}
\usepackage{graphicx} 
\usepackage{epstopdf}
\usepackage{graphicx,epstopdf,color}
\usepackage{amsfonts}
\usepackage{amsmath,amssymb,mathrsfs}
\usepackage{subfigure}
\usepackage{hyperref}
\usepackage{bm}

\def\ea{\emph{et al.}}

\allowdisplaybreaks[1]
\begin{document}

\title{Correlated Dirac particles and Superconductivity on the Honeycomb Lattice}

\author{Wei Wu}
\affiliation{D\' epartement de Physique and Regroupement Qu\' eb\' ecois sur les Mat\' eriaux de Pointe, Universit\' e de Sherbrooke, Sherbrooke, QC J1K 2R1, Canada}
\affiliation{Department of Physics, Yale University, New Haven, CT 06520, USA}
\author{Michael  M. Scherer}
\affiliation{Institute for Theoretical Physics, University of Heidelberg, D-69120 Heidelberg, Germany}
\affiliation{Institute for Theoretical Solid State Physics, RWTH Aachen University, D-52056 Aachen, Germany \\ and JARA Fundamentals of Future Information Technologies}
\author{Carsten Honerkamp}
\affiliation{Institute for Theoretical Solid State Physics, RWTH Aachen University, D-52056 Aachen, Germany \\ and JARA Fundamentals of Future Information Technologies}  
\author{Karyn Le Hur}
\affiliation{Centre de Physique Th\' eorique, \' Ecole Polytechnique, CNRS, 91128 Palaiseau C\' edex, France}
\affiliation{Department of Physics, Yale University, New Haven, CT 06520, USA}

\begin{abstract}
We investigate the properties of  the nearest-neighbor singlet pairing and the emergence of d-wave superconductivity in the doped honeycomb lattice considering the limit of large interactions and the $t-J_1-J_2$ model.  First, by applying a renormalized mean-field procedure as well as slave-boson theories which account for the proximity to the  Mott insulating state,  we confirm the emergence of d-wave superconductivity in agreement with earlier works. We show that a small but finite $J_2$ spin coupling between next-nearest neighbors stabilizes d-wave symmetry compared to the extended s-wave scenario. At small hole doping, to minimize energy and to gap the whole Fermi surface or all the Dirac points, the superconducting ground state is characterized by a $d+id$ singlet pairing assigned to one valley and a $d-id$ singlet pairing to the other, which then preserves time-reversal symmetry. The slightly doped situation is distinct  from the heavily doped case (around 3/8 and 5/8 filling) supporting a pure chiral $d+id$ symmetry and breaking  time-reversal symmetry. Then, we apply the functional Renormalization Group and we study in more detail the competition between antiferromagnetism and superconductivity in the vicinity of half-filling. We discuss possible applications to strongly-correlated compounds with Copper hexagonal planes such as In$_3$Cu$_{2}$VO$_9$. Our findings are also relevant to the understanding of exotic superfluidity with cold atoms.
\end{abstract}

\maketitle

\section{Introduction}

Recently, graphene systems have attracted a considerable attention of experimentalists as well as theorists \cite{RMP_graphene,eva_graphene}. Graphene which consists of a single layer of carbon atoms forming a honeycomb lattice allows to realize in a condensed-matter system the Dirac equation, where electrons behave as massless Dirac fermions. The observation of massless Dirac fermions in monolayer graphene has engendered a new era of science and electrons in graphene embody a typical example of weakly interacting relativistic quantum systems \cite{Novoselov,Kim}. The chemical potential can be tuned and hence it is possible to change the concentration of carriers, holes or electrons, opening the door for carbon based electronics. It is also relevant to mention that artificial graphene has also been realized in cold atom systems \cite{Tarruell}, with photons \cite{Houck,AlexKaryn,new} and using scanning tunneling microscopy techniques \cite{Hari}. Doped graphene exhibits a finite density of states which favors antiferromagnetic spin fluctuations \cite{AFM} and then may lead to unconventional superconductivity. Experimentally, superconductivity in graphene has been induced by proximity effect through contact with superconducting electrodes \cite{SCprox}. This shows that Cooper pairs can propagate coherently in graphene.

Several theoretical attempts have been made to describe the emergence of superconductivity in graphene 
\cite{Uchoa,BlackSchaffer,Honerkamp,Baskaran,NCL} as well as the formation of zero-energy states in the cores of vortices or at the boundaries \cite{Ghaemi,DoronKaryn,Black2}. In general, if these bound states appear in even number then they are not topologically protected and, for example, for small coherence lengths \cite{DoronKaryn} their energies approach the Caroli, De Gennes, Matricon energy \cite{CGM}. Uchoa {\it et al.} \cite{Uchoa} suggested that an extended s-wave SC phase may be realized at the mean-field level due to the peculiar structure of the honeycomb lattice. On the other hand, for a purely on-site repulsive Hubbard interaction $U$, as shown through the functional Renormalization Group (fRG) \cite{Honerkamp}, the nearest-neighbor spin exchange interaction $J$ can lead to a $d+id$ $(d_{xy}+ id_{x^2-y^2})$ superconducting state as a reminiscence of the superconductivity on the triangular lattice \cite{Lee_p}. A similar fRG scheme has been applied on the square lattice \cite{ZS,Hallboth,CarstenMaurice,CarstenManfred}. The $d+id$ superconducting state has also been found using a mean-field theory on a toy model with singlet pairing between different sublattices \cite{BlackSchaffer,Baskaran}. A similar result has also been confirmed via numerical results based on a recently developed variational method, the Grassmann tensor product state approach \cite{Gu}. These theoretical investigations concern the situation close to half-filling. 

On the other hand, experimental techniques in doping methods \cite{dope1,dope2} have allowed to approach the van Hove singularities, which corresponds to dope the graphene close to the M point of the Brillouin zone, {\it i.e.}, for 3/8 or 5/8 electron filling (which corresponds to doping $\pm 1/8$ from the Dirac points; pristine graphene corresponds to 1/2 filling). The logarithmically divergent density of states at the van Hove singularities (van Hove filling) unambiguously favors the appearance of $d+id$ superconductivity for weak repulsive on-site interactions, as shown from a perturbative RG approach \cite{NCL} and a fRG framework \cite{Kiesel,Wang}. Rather unique on the honeycomb lattice is the degeneracy of the two d-wave pairing channels \cite{BlackSchaffer,Gonzalez}.

In graphene, the Hubbard interaction is approximately half the bandwidth which places this material in the intermediate-coupling regime. Below, motivated by the recent realization of strongly-correlated honeycomb lattice materials such as In$_3$Cu$_2$VO$_9$ \cite{Kataev,Mydosh,Yan}, we are rather interested in the stronger interaction regime which allows to realize Mott physics in the half-filled situation. A similar situation could be eventually reached using cold atomic systems on the honeycomb lattice \cite{Tarruell}. The honeycomb lattice, which is a bipartite lattice, allows for a spin density wave order \cite{Sorella,Herbut,Wei,RachelLeHur}. In In$_3$Cu$_2$VO$_9$, singly occupied $3z^2-r^2$ electrons of coppers contribute to an antiferromagnetic moment with $S=1/2$. At quite intermediate values of the Hubbard interaction, using Quantum Monte Carlo (QMC) simulations and an accurate finite-size scaling up to 648 sites, Meng {\it et al.} \cite{Meng} have recently reported the emergence of a spin liquid ground state in the range $3.5\leq U/t\leq 4.3$, which is characterized by a single-particle gap where band theory would predict a metallic behavior; see also Refs. \onlinecite{Hohenadler,CWu}. The fingerprints of such a Mott phase without long-range N\' eel ordering have also been reported using cluster methods \cite{Wu,Li} when increasing the size of the cluster unit cell \cite{Wu} and in anisotropic lattices \cite{Wangetal}. Sorella {\it et al.} \cite{Sorella2} extended the QMC calculations up to 2592 sites and  did not find evidence for this spin liquid phase region. On the other hand, the existence of a spin liquid phase, with a spin gap and ${\cal Z}_2$ symmetry, has been corroborated in the strong-coupling $J_1-J_2$ effective spin model on the honeycomb and square lattices \cite{Clark,Ran,Balents}. For large values of $J_2$, one may also expect a dimerized symmetry-broken phase \cite{Clark,Lhuillier}. On the honeycomb lattice, in particular, this invalidates the possibility of an algebraic spin liquid  with U(1) or SU(2) gauge theories at relatively moderate interactions  \cite{LeeLee,Hermele}. On the other hand, stable algebraic spin liquids on the honeycomb lattice with $Z_2$ symmetry have been predicted for strong interactions \cite{Kitaev,Senechal,Ruegg}. The undoped compound In$_3$Cu$_2$VO$_9$ seems to yield a magnetically ordered ground state \cite{Mydosh,Yan}. Finding a spin liquid in two or three dimensions represents a considerable challenge in condensed-matter physics \cite{Andersonspinliquid,Doucot,Kivelson,Read2,Wen2,Wen0,Balents2,Misguich,Lee,Subir,xu}. It is also relevant to note that a spin liquid ground state has also been  reported in two-dimensional Kagom\' e \cite{Mendels,Fak} and organic triangular materials \cite{un,deux,trois,quatre} in relation with theoretical developments \cite{LeeLee,Kato,Kitaev,Misguich,Lee,Messio,Balents3,Guo,Santos,Isakov,Wen2,Fiete,Melko,Balents2, HuseWhite,Schollwoeck,Poilblanc}. 

In this paper, we seek to start from the Mott insulating and N\' eel ordered phase on the honeycomb lattice and dope the system away from half-filling with a few holes. Our main goal is to investigate the emergence of pairing and superconductivity within the framework of the $t-J_1-J_2$ model, when including a finite (but small) next nearest-neighbor spin exchange interaction. A strong correlation view of the Hubbard model, through the $t-J$ model \cite{tJ}, was advanced by Anderson \cite{Andersonspinliquid}, who conjectured the relevance of a spin liquid phase or Resonating Valence Bond (RVB) phase as a result of the motion of the holes, destroying the antiferromagnetic order. The RVB state corresponds to a spin-gapped singlet state with no symmetry breaking. The doped spin-1/2 honeycomb lattice compound In$_3$Cu$_2$VO$_9$ might be a good candidate for the realization of such a physics through the $t-J_1-J_2$ model \cite{Yan2}. The condensation of the holes (bosons) at low temperatures should result in a superconducting ground state. 

On the square lattice, following the Gutzwiller projector point of view \cite{Gutzwiller,BrinkmanRice}, this scenario has been pushed forward through a projected mean-field theory (the renormalized mean-field theory or Gutzwiller RVB theory) removing all components of the wavefunction with doubly occupied sites \cite{Gros,Anderson2,Gros2,Ogata,LeHurRice,Rice,ALFK}, ``slave-particle'' approaches \cite{Barnes,Read,Coleman,KotliarRuckenstein,Kotliar1,Affleck,Baskaran2,WenL,Woelfle,Florens,WenLee,Biermann,Pepin,Ruegg,Senthil} and powerful numerical approaches \cite{GiamarchiLhuillier,Paramekanti,Georges,Vollhardt2,Jarrell,Tremblay,Kotliar,Gull,Kozik,Sordi,Millis}. We shall also mention some theoretical progess accomplished close to the Mott state \cite{LeHurRice,Tesanovic,RMPK,Pines,Konik,CaseyAnderson,Shastry}. 

By applying the renormalized mean-field theory or Gutzwiller RVB theory \cite{Gros,Anderson2,Gros2,Ogata,LeHurRice,Rice} and ``slave-particle'' approaches  \cite{Barnes,Read,Coleman,KotliarRuckenstein,Kotliar1,Affleck,Baskaran2,WenL,Woelfle,Florens,WenLee,Biermann,Pepin,Ruegg,Senthil}, first we will show that on the honeycomb lattice and close to half-filling, the $d\pm id$ pairing order parameter favoring spin singlet between nearest neighbors is stabilized by the small $J_2$ antiferromagnetic spin exchange whereas the extended s-wave pairing strength diminishes. For strong interactions, as noticed in Ref. \onlinecite{Spin}, the ground state taking $d+id$ in one valley and $d-id$ in the other does not break time-reversal symmetry in contrast to the heavily-doped situation at 3/8 or 5/8 electron filling at weak interactions \cite{NCL,Kiesel,Wang} and minimizes the total energy since the whole Fermi surface becomes gapped. The edge state from the $d+id$ pairing in one valley is canceled by that from the $d-id$ pairing in the other valley. This assignment turns out to be essential because the $d+id$ order parameter in one valley vanishes in the other valley, allowing the Dirac spectrum to be gapless \cite{Spin}. Using the fRG approach, then we allow for the presence of antiferromagnetism at half-filling and study more rigorously the competition between superconductivity and antiferromagnetism in the presence of the $J_2$ term, following the scheme of Ref. \onlinecite{Honerkamp}.

The remainder of the paper is organized as follows. In Sec. II, we introduce the model Hamiltonian, discuss the dominant pairing symmetries using the renormalized mean-field theory (RMFT) and the effect of a finite next-nearest neighbor spin exchange $J_2$. We also comment on the possibility of stable ${\cal Z}_2$ (gapped) spin liquids at half-filling for not too small values of $J_2$. In Sec. III, we present the theoretical framework, the main equations and the results. In Sec. IV, by applying fRG, we address the competition between antiferromagnetism and superconductivity as a function of $J_2$ and doping. In Appendix A, we compare our results obtained from the RMFT with those obtained within the U(1) slave-boson theory.

\section{Model Hamiltonian}

To capture the effect of strong interactions (or Mott physics at half-filling) in honeycomb lattice compounds such as In$_3$Cu$_{2}$VO$_9$ \cite{Kataev,Mydosh,Yan}, we consider the (renormalized) $t-J_{1}-J_{2}$ model:
\begin{eqnarray}
H &=& -tg_{t}\sum_{\langle i,j\rangle\sigma}\left(c_{i\sigma}^{\dagger}d_{j\sigma}+h.c.\right) \\ \nonumber
&-&\mu\sum_{i\sigma}\left(c_{i\sigma}^{\dagger}c_{i\sigma}+d_{i\sigma}^{\dagger}d_{i\sigma}\right)
\\ \nonumber
 & +& J_1 g_{1}\sum_{\langle i,j\rangle}{\bf S}_{i}\cdot {\bf S}_{j}+J_{2}g_{2}\sum_{\langle\langle i,j\rangle\rangle}{\bf S}_{i}\cdot {\bf S}_{j}.
\end{eqnarray}
The Gutzwiller projector \cite{Gutzwiller} ensuring that configurations with doubly occupied sites are forbidden is replaced by statistical weighting factors, $g_{t}=2\delta/(1+\delta)$ \cite{Vollhardt} and $g_{1}=g_{2}=4/(1+\delta)^{2}$ \cite{Gros}, and explicitly depend on the doping level $\delta$ 
\cite{Anderson2,Gros,Gros2,Ogata,LeHurRice,Rice}. Note that here, $\delta=1-n$ where $n$ refers to the number of electrons per site. For the half-filled situation, the number of holes per site is $\delta=0$. In addition, $\langle i,j\rangle$ denotes a nearest neighbor pair where  $i<j$ is assumed. We also denote $c$ and $d$ electron annihilation operators associated with the two sublattices ($A$ and $B$) of the honeycomb lattice. 

In the limit of large on-site interaction, this model can be derived from the Hubbard model, similarly to the derivation of the Kondo model from the Anderson model \cite{SchriefferWolff}, by resorting to a perturbation theory in $t/U$ up to fourth order processes \cite{Clark,Q}
\begin{equation}
J_1 = \frac{4t^2}{U} - \frac{16 t^4}{U^3}, \hskip 1cm J_2=\frac{4t^4}{U^3}.
\end{equation}
Note that an antiferromagnetic phase has been reported for $J_2/J_1<0.08$ \cite{Clark} which corresponds to $U/t\approx 4.3$. This is in agreement with QMC results of Refs. \onlinecite{Meng,Hohenadler} and Cluster methods of Refs. \onlinecite{Wu,Li}, but in contrast with the very recent QMC results found in Ref. \onlinecite{Sorella2} which predict that the spin density wave order on the honeycomb lattice would appear simultaneously with the Mott transition at lower interaction strength. 

Below, we start from half-filling with the spin density wave order where $J_2<0.08 J_1$. Note that for the undoped compound In$_3$Cu$_{2}$VO$_9$, it has been recently estimated that $J_2/J_1\approx 0.04$ \cite{Yan2}. The undoped compound seems to order at low temperatures  \cite{Kataev,Mydosh,Yan}. Hereafter, we do not focus on the half-filled situation and assume that there is a finite hole concentration. By increasing the number of carriers, one may expect an RVB type scenario and a gapped spin liquid \cite{Andersonspinliquid}, as a result of the motion of the carriers (holes). Hereafter, we describe this aspect of the problem through the $t-J_1-J_2$ model close to half-filling.

\subsection{Pairing symmetries}

Firstly, we investigate the t-J model where $J_1=J$ and $J_{2}=0$ applying the RMFT  \cite{Gros,Anderson2,Gros2,Ogata,LeHurRice,Rice,LeHurPaul}. The emergence of nearest neighbor singlet pairing for repulsive on-site interactions on the honeycomb lattice has been first discussed in Refs. \onlinecite{BlackSchaffer,Honerkamp}. Our procedure is slightly different from the one used by Black-Schaffer and Doniach \cite{BlackSchaffer} since we take into account the large interaction limit through the statistical weighting factors $g_t$, $g_1$ and $g_2$, and we shall also address the effect of a finite next nearest neighbor coupling $J_2$.

Following the procedure used on the square lattice \cite{Anderson2,Gros,Gros2,Ogata,LeHurRice,Rice}, it is convenient to introduce the mean-field order parameters,
\begin{eqnarray}
\begin{split}\chi_{ij} & =\frac{3}{4}g_{1}J \sum_{\sigma}\langle c_{i\sigma}^{\dagger}d_{j\sigma}\rangle\\
\Delta_{ij} & =\frac{3}{4}g_{1}J \langle c_{i\uparrow}d_{j\downarrow}-c_{i\downarrow}d_{j\uparrow}\rangle,
\end{split}
\end{eqnarray}
and we focus on the nearest-neighbor singlet pairing contribution (on-site pairing is forbidden due to the very large on-site repulsion). 
Assuming the uniform solution for the $\chi$ field the mean-field Hamiltonian takes the form,
\begin{eqnarray}
\label{pairing}
H & = & \left(-tg_{t}-\frac{\chi}{2}\right)\sum_{\langle ij\rangle\sigma}\left(c_{i\sigma}^{\dagger}d_{j\sigma}+d_{j\sigma}^{\dagger}c_{i\sigma}\right)\\ \nonumber
&- & \frac{1}{2}\sum_{\langle ij\rangle}\left(\Delta_{ij}\left(c_{i\uparrow}^{\dagger}d_{j\downarrow}^{\dagger}+d_{j\uparrow}^{\dagger}c_{i\downarrow}^{\dagger}\right)+h.c.\right)\\ \nonumber
 & +&\frac{1}{3}\sum_{\langle ij\rangle}\frac{|\chi|^{2}}{{J g_{1}}}+\frac{1}{3}\sum_{\langle ij\rangle}\frac{|\Delta_{ij}|^{2}}{J g_{1}}\\ \nonumber
 & -&\left(\mu-\frac{J g_{1}}{4}\right)\sum_{i\sigma}\left(c_{i\sigma}^{\dagger}c_{i\sigma}+d_{i\sigma}^{\dagger}d_{i\sigma}\right).
\end{eqnarray}
Here, we have released the constraint $i<j$ and introduced the chemical potential $\mu$ explicitly. It is judicious to Fourier transform the Hamiltonian and introduce the symmetric and antisymmetric (band) combinations of the electron operators $c$ and $d$ which diagonalize the kinetic part, as follows
\begin{eqnarray}
c_{{\bf k} \sigma} &=& \frac{1}{\sqrt{2}}(f_{{\bf k}\sigma} +g_{{\bf k}\sigma}) \\ \nonumber
d_{{\bf k}\sigma} &=& \frac{1}{\sqrt{2}}\exp(-i\phi_{\bf k})(f_{{\bf k}\sigma} - g_{{\bf k}\sigma}).
\end{eqnarray}
This results in the Hamiltonian:
\begin{eqnarray}
H &=& \sum_{k\sigma}\left(\epsilon_{{\bf k}}-\mu\right)f_{{\bf k}\sigma}^{\dagger}f_{{\bf k}\sigma}+\sum_{{\bf k}\sigma}(-\epsilon_{{\bf k}}-\mu)g_{{\bf k}\sigma}^{\dagger}g_{{\bf k}\sigma}\nonumber \\
 &-& \sum_{{\bf k}}\left(\Delta_{{\bf k}}^{i}\left(f_{{\bf k}\uparrow}^{\dagger}f_{-{\bf k}\downarrow}^{\dagger}-g_{{\bf k}\uparrow}^{\dagger}g_{-{\bf k}\downarrow}^{\dagger}\right)+h.c.\right) \nonumber \\
 &+& \sum_{{\bf k}}\left(\Delta_{{\bf k}}^{I}\left(f_{{\bf k}\uparrow}^{\dagger}g_{-{\bf k}\downarrow}^{\dagger}-g_{{\bf k}\uparrow}^{\dagger}f_{-{\bf k}\downarrow}^{\dagger}\right)+h.c.\right) \nonumber \\
&+& \frac{N_{s}|\chi|^{2}}{J g_{1}}+\frac{N_{s}}{3}\frac{\sum_{\alpha}|\Delta_{\alpha}|^2}{J g_{1}}.
\end{eqnarray}
Hereafter, the sum over $\alpha$ corresponds to a summation over the three nearest neighbors on the honeycomb lattice and $\Delta_{\alpha}$ is defined in Eq. (7). Here, $N_s$ corresponds to the total number of sites, $\Delta_{{\bf k}}^{i}$ is the intraband pairing while $\Delta_{{\bf k}}^{I}$ is the interband counterpart breaking time reversal symmetry,
\begin{eqnarray}
\Delta_{{\bf k}}^{i} &=& \frac{1}{2} \sum_{\alpha}\Delta_{\alpha}\cos({\bf k}\cdot {\bf R}_{\alpha}-\phi_{{\bf k}})\\ \nonumber
\Delta_{{\bf k}}^{I} &=& \frac{1}{2}\sum_{\alpha}\Delta_{\alpha}i \sin({\bf k}\cdot {\bf R}_{\alpha}-\phi_{{\bf k}})\\ \nonumber
\epsilon_{{\bf k}} & =& \left(-tg_{t}-\frac{\chi}{2}\right)|\gamma_{{\bf k}}|.
\end{eqnarray}
It is convenient to define $\gamma_{{\bf k}} = \sum_{\alpha}e^{i {\bf k}\cdot {\bf R}_{\alpha}}$. The phase $\phi_{\bf k}$ is defined as $\phi_{{\bf k}}=\hbox{arg}(\sum_{\alpha}e^{i{\bf k}\cdot {\bf R}_{\alpha}})=-\phi_{-{\bf k}}$ and satisfies the following relation $\exp(i\phi_{\bf k})\gamma_{\bf k} = \exp(-i\phi_{\bf k})\gamma_{\bf k}^*=|\gamma_{\bf k}|$. 

At a general level, the intraband pairing contribution exhibits an order parameter {\it even} in {\bf k} space, and one can check that it corresponds to the {\it singlet} pairing state. In contrast, the interband pairing contribution has an order parameter {\it odd} in {\bf k} space. For a bond-independent s-wave order parameter, the interband (spinon) pairing then is identically zero, but this is not necessarily the case for an arbitrary wave symmetry (such as d-wave symmetry). In this paper, we mostly consider the two dominant nearest-neighbor singlet pairing order parameters when assuming purely on-site interactions, namely the extended s-wave pairing (ES) and the $d\pm id$ pairing \cite{Honerkamp,BlackSchaffer}. For the ES pairing only the intraband pairing form factor $\Delta_{{\bf k}}^{i}$ are non-zero, 
\begin{eqnarray}
\Delta_{{\bf k}}^{i} & =& \frac{1}{2}\sum_{\alpha}\Delta e^{i{\bf k}\cdot {\bf R}_{\alpha}}\\ \nonumber
\Delta_{{\bf k}}^{I} & =& 0.
\end{eqnarray}
For the $d\pm id$ pairing state, the order parameter can be viewed as a mixture of $d_{xy}$ and $d_{x^2-y^2}$ \cite{Honerkamp,BlackSchaffer}:
\begin{equation}
\Delta_{{\bf k}}^{d\pm id}=\cos\left(\frac{\pi}{3}\right)\Delta_{x^{2}-y^{2}}({\bf k})\pm i\sin\left(\frac{\pi}{3}\right)\Delta_{xy}({\bf k}). 
\end{equation}
In fact, it is perhaps judicious to remember that the $d_{x^2-y^2}$ and $d_{xy}$ wave symmetry functions satisfy:
\begin{eqnarray}
d_{x^2-y^2}(k_x,k_y) &=& e^{-i\frac{k_x}{\sqrt{3}}} - e^{i\frac{k_x}{2\sqrt{3}}}\cos\left(\frac{k_y}{2}\right) \\ \nonumber
d_{xy}(k_x,k_y) &=& ie^{i\frac{k_x}{2\sqrt{3}}} \sin\left(\frac{k_y}{2}\right).
\end{eqnarray}
Therefore, it is convenient to introduce the notations:
\begin{eqnarray}
\Delta_{{\bf k}}^{i} &=& \frac{1}{2}\sum_{\alpha}\Delta_{\alpha}\cos({\bf k}\cdot {\bf R}_{\alpha}-\phi_{\bf k})=\Delta\Gamma_{{\bf k}}^{i}\\ \nonumber
\Gamma_{{\bf k}}^{i} &=& \frac{1}{2}\sum_{\alpha}e^{2i\pi(\alpha-1)/3}\cos({\bf k}\cdot {\bf R}_{\alpha}-\phi_{{\bf k}})\\ \nonumber
\Delta_{{\bf k}}^{I} &=& \frac{1}{2}\sum_{\alpha}\Delta_{\alpha}i \sin({\bf k}\cdot {\bf R}_{\alpha}-\phi_{{\bf k}})=\Delta\Gamma_{{\bf k}}^{I}\\ \nonumber
\Gamma_{{\bf k}}^{I}&=&\frac{1}{2}\sum_{\alpha}e^{2i\pi(\alpha-1)/3}i\sin({\bf k}\cdot {\bf R}_{\alpha}-\phi_{{\bf k}}).
\end{eqnarray}

In fact, owing to the large overlap between the nodes of the ES form factor and the Fermi surface (see Fig. 1), one could anticipate that the ES solution will have a higher free energy then favoring the $d\pm id$ pairing symmetry. 
The mean-field equations will be discussed in the next Sec. III.

\begin{figure}
\includegraphics[scale=0.45]{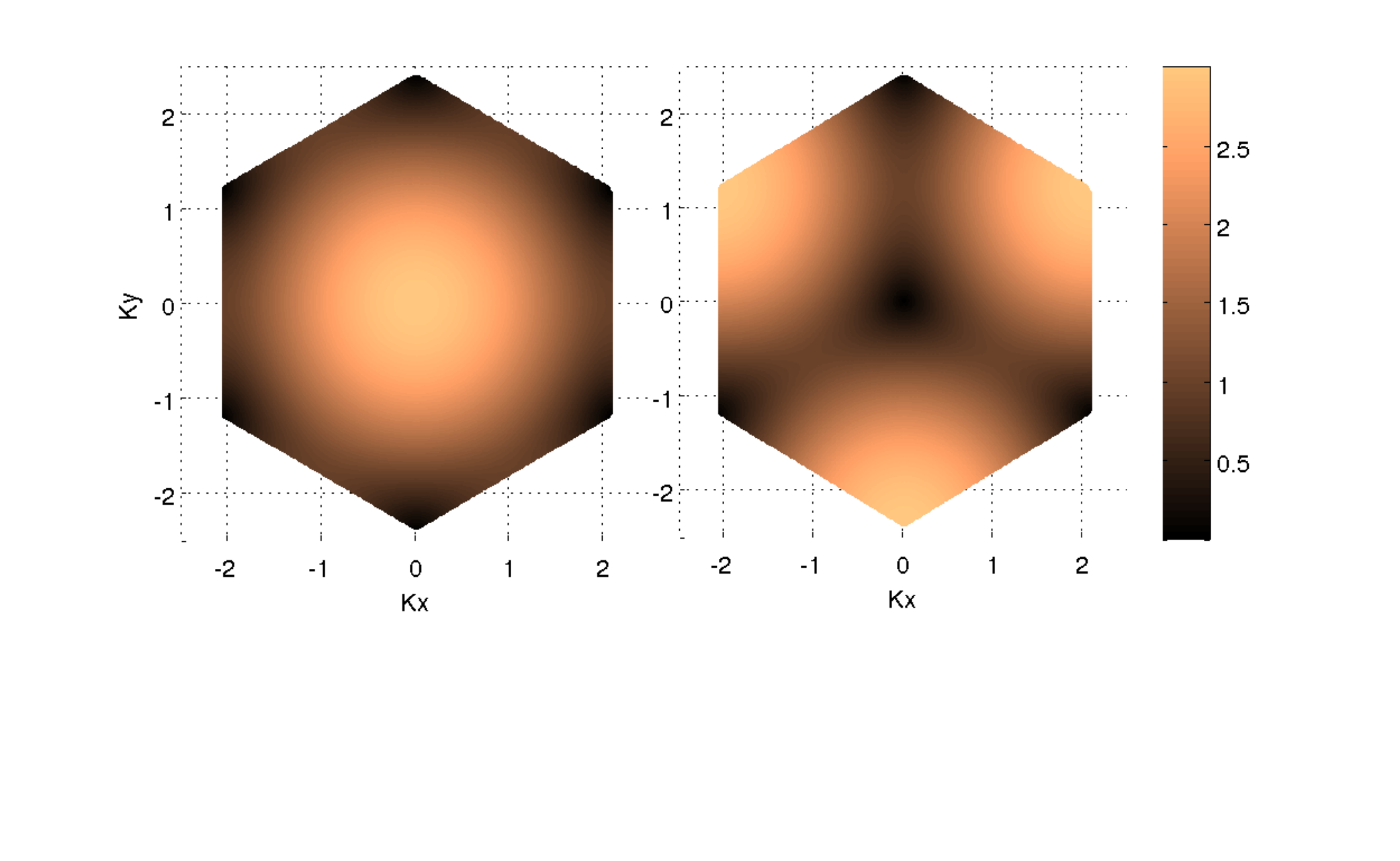}
\vskip -1.4cm
\caption{Form factors of the ES and $d+id$ pairing solutions. Due to the overlap between the Fermi surface and the nodes of the ES solution, the ground state should favor the $d+id$ pairing as discussed in Sec. III. On the other hand, to gap all the Fermi surface close to half-filling, the solution that minimizes the whole energy will be taking the $d+id$ pairing solution in one valley and the $d-id$ in the other \cite{Spin}.}
\end{figure}

\subsection{Effect of $J_2$}

To investigate the effect of the next nearest neighbor spin exchange $J_2$ on the physical properties of the system, here we will assume that since we consider the limit where $J_1\gg J_2$, the dominant pairing order parameter is $\Delta_{ij}$ which corresponds to (d-wave symmetry for) nearest neighbor singlet pairing. As a result, we only introduce an extra particle-hole order parameter which couples the next nearest neighbor sites:
\begin{eqnarray}
\chi' & = & \frac{3}{4}g_{2}J_{2}\sum_{\sigma}\langle c_{i\sigma}^{\dagger}c_{j\sigma}\rangle.
\end{eqnarray}
The mean-field Hamiltonian on the honeycomb lattice then takes the following form:
\begin{eqnarray}
H & =& \left(-tg_{t}-\frac{\chi}{2}\right)\sum_{\langle ij\rangle\sigma}\{c_{i\sigma}^{\dagger}d_{j\sigma}+d_{j\sigma}^{\dagger}c_{i\sigma}\} \\ \nonumber
&-&\frac{\chi'}{2}\sum_{\ll ij\gg\sigma}\{c_{i\sigma}^{\dagger}c_{j\sigma}+d_{i\sigma}^{\dagger}d_{j\sigma}\}\\ \nonumber
&- & \frac{1}{2}\sum_{\langle ij\rangle}\large{\Delta_{ij}\left(c_{i\uparrow}^{\dagger}d_{j\downarrow}^{\dagger}+d_{j\uparrow}^{\dagger}c_{i\downarrow}^{\dagger}\right)+h.c.\large}\\ \nonumber
 &+& \frac{1}{3}\sum_{\langle ij\rangle}\frac{|\chi|^{2}}{J_{1}g_{1}}+\frac{1}{3}\sum_{\langle ij\rangle}\frac{|\Delta_{ij}|^{2}}{J_{1}g_{1}}
 +\frac{N_{s}z'}{3}\frac{|\chi'|^{2}}{J_{2}g_{2}}\\ \nonumber
 & -& \mu\sum_{i\sigma}\left(c_{i\sigma}^{\dagger}c_{i\sigma}+d_{i\sigma}^{\dagger}d_{i\sigma}\right).
\end{eqnarray}
Here, $z'$ denotes the next nearest-neighbor coordination number, i.e., $z'=4$ for the square lattice and $z'=6$ for the honeycomb lattice. After Fourier transformation, we obtain:
\begin{eqnarray}
H &=& \left(-tg_{t}-\frac{\chi}{2}\right)\sum_{{\bf k}\sigma}\{\gamma_{{\bf k}}c_{{\bf k}\sigma}^{\dagger}d_{{\bf k}\sigma}+\gamma_{{\bf k}}^{*}d_{{\bf k}\sigma}^{\dagger}c_{{\bf k}\sigma}\} \\ \nonumber
&-&\frac{\chi'}{2}\sum_{{\bf k}\sigma}\{\zeta_{{\bf k}}c_{{\bf k}\sigma}^{\dagger}c_{{\bf k}\sigma}+\zeta_{{\bf k}}^{*}d_{{\bf k}\sigma}^{\dagger}d_{{\bf k}\sigma}\}\\ \nonumber
&-& \frac{1}{2}\sum_{{\bf k}}\{\Delta_{{\bf k}}(c_{{\bf k}\uparrow}^{\dagger}d_{-{\bf k}\downarrow}^{\dagger}-c_{{\bf k}\downarrow}^{\dagger}d_{-{\bf k}\uparrow}^{\dagger})+h.c.\}\\ \nonumber
 &+& \frac{1}{3}\sum_{\langle ij\rangle}\frac{|\chi|^{2}}{J_{1}g_{1}}+\frac{1}{3}\sum_{\langle ij\rangle}\frac{|\Delta_{ij}|^{2}}{J_{1}g_{1}}+\frac{N_{s}z'}{3}\frac{|\chi'|^{2}}{J_{2}g_{2}}\\ \nonumber
 & -& \mu\sum_{{\bf k}\sigma}\{c_{{\bf k}\sigma}^{\dagger}c_{{\bf k}\sigma}+d_{{\bf k}\sigma}^{\dagger}d_{{\bf k}\sigma}\}
\end{eqnarray}
where,
\begin{eqnarray}
\gamma_{{\bf k}} & =& \sum_{\alpha}e^{i{\bf k}\cdot {\bf R}_{\alpha}}\\ \nonumber
\zeta_{{\bf k}} & =& \zeta_{{\bf k}}^{*}=\sum_{\beta}e^{i {\bf k}\cdot {\bf R}_{\beta}}\\
\Delta_{{\bf k}} & =& \sum_{\alpha}\Delta_{\alpha}e^{i{\bf k}\cdot {\bf R}_{\alpha}}.
\end{eqnarray}
The sum over $\beta$ here denotes the summation over next nearest neighbors. 

Assuming that $J_1\gg J_2$, we then have a modified dispersion relation for the fermions $\xi_{{\bf k}}=(-tg_{t}-\frac{\chi}{2})|\gamma_{{\bf k}}|-\frac{\chi'}{2}\zeta_{{\bf k}}$ and an additional constant term $\frac{2N_{s}|\chi'|^{2}}{J_{2}g_{2}}$. The self-consistent equations for the $t-J_{1}-J_{2}$ model then will be directly inferred from the ones for the $t-J$ model. However, we shall have an additional equation determining $\chi'$.  In Sec. III, we shall show that the $J_2$ term through the extra order parameter $\chi'$ will help stabilizing the $d\pm id$ spin pairing.

\section{Results from RMFT}

Here, we give the main equations associated with the different pairing solutions by resorting to the renormalized mean-field theory.

\subsection{Extended s-wave scenario}

As discussed earlier in Sec. II, for the ES pairing, only the intraband pairing form factors of Eq. (9) are non-zero. Since the effect of $J_2$ is relatively simple following the scheme above, we present the main equations for $J_2=0$. After standard Bogoliubov transformation, the Hamiltonian can be formally re-written as:
\begin{eqnarray}
H &=& E_{0}+\sum_{{\bf k}l}E_{{\bf k}l}\{a_{{\bf k}l}^{\dagger}a_{{\bf k}l}+a_{-{\bf k}l}^{\dagger}a_{-{\bf k}l}\} \\ \nonumber
&+& \frac{N_{s}|\chi|^{2}}{J_{1}g_{1}}+\frac{N_{s}|\Delta|^{2}}{J_{1}g_{1}}.
\end{eqnarray}
Here, the sum $l=0,1$ stems from the path integral of the two-band Hamiltonian and
\begin{equation}
\begin{split}
E_{0} & =\sum_{{\bf k}l}\{(-1)^{l}\xi_{{\bf k}}-\mu\}-\sum_{{\bf k}l}E_{{\bf k}l}\\
E_{{\bf k}l} & =\sqrt{\{(-1)^{l}\xi_{{\bf k}}-\mu\}^{2}+\frac{1}{4}|\Delta_{{\bf k}}|^{2}},
\end{split}
\end{equation}
where we have introduced $E_{{\bf k}l}=\sqrt{\xi_{{\bf k}l}^{2}+\frac{1}{4}|\Delta_{{\bf k}}|^{2}}$ with $\Delta_{{\bf k}} =\sum_{\alpha}\Delta e^{i{\bf k}\cdot {\bf R}_{\alpha}}$ and $\xi_{{\bf k}l}=(-1)^{l}(-tg_{t}-\frac{\chi}{2})|\gamma_{{\bf k}}|-\mu$ when $J_2=0$. 
The free energy then takes the form
\begin{eqnarray}
F &=& -2T\sum_{{\bf k}l}\ln\left(2\cosh\frac{\beta E_{{\bf k}l}}{2}\right) \\ \nonumber
&-& N_{s}\mu+\frac{N_{s}|\chi|^{2}}{J g_{1}}+\frac{N_{s}|\Delta|^{2}}{Jg_{1}}.
\end{eqnarray}
For simplicity, we set the Boltzmann constant $k_B=1$. At the stationary point of the free energy $F$ we obtain the BCS-like self-consistent equations,
\begin{eqnarray}
\delta & =& \frac{1}{N_{s}}\sum_{{\bf k}l}\frac{\xi_{{\bf k}l}}{E_{{\bf k}l}}\tanh\frac{\beta E_{{\bf k}l}}{2}\\ \nonumber
\chi & =& -\frac{Jg_{1}}{4N_{s}}\sum_{{\bf k}l}\frac{(-1)^{l}\xi_{{\bf k}l}|\gamma_{{\bf k}}|}{E_{{\bf k}l}}\tanh\frac{\beta E_{{\bf k}l}}{2}\\ \nonumber
\Delta &=& \frac{Jg_{1}}{8N_{s}}\sum_{{\bf k}l}\frac{\Delta|\gamma_{{\bf k}}|^{2}}{E_{{\bf k}l}}\tanh\frac{\beta E_{{\bf k}l}}{2}.
\end{eqnarray}
The solution of these equations will be discussed in Sec. III. C. Now, we present the equations for the $d\pm id$ case.

\subsection{d-wave scenario}

For the $d\pm id$ case, we can rewrite the Hamiltonian by introducing $\Phi_{{\bf k}}^{\dagger}=[f_{{\bf k}\uparrow}^{\dagger},f_{-{\bf k}\downarrow},g_{{\bf k}\uparrow}^{\dagger},g_{-{\bf k}\downarrow}]$, 

\begin{eqnarray*}
H & =\sum_{{\bf k}}\Phi_{{\bf k}}^{\dagger}\begin{bmatrix}\xi_{{\bf k}0} & -\Delta_{{\bf k}}^{i} & 0 & \Delta_{{\bf k}}^{I}\\
-\Delta_{{\bf k}}^{i*} & -\xi_{{\bf k}0} & -\Delta_{{\bf k}}^{I*} & 0\\
0 & -\Delta_{{\bf k}}^{I} & \xi_{{\bf k}1} & \Delta_{{\bf k}}^{i}\\
\Delta_{{\bf k}}^{I*} & 0 & \Delta_{{\bf k}}^{i*} & -\xi_{{\bf k}1}
\end{bmatrix} & \Phi_{{\bf k}}+\hbox{Const}.
\end{eqnarray*}

\begin{widetext}
Thus we can determine the energy dispersion of the Bogoliubov quasiparticles,
\begin{figure}[ht]
\includegraphics[scale=0.37]{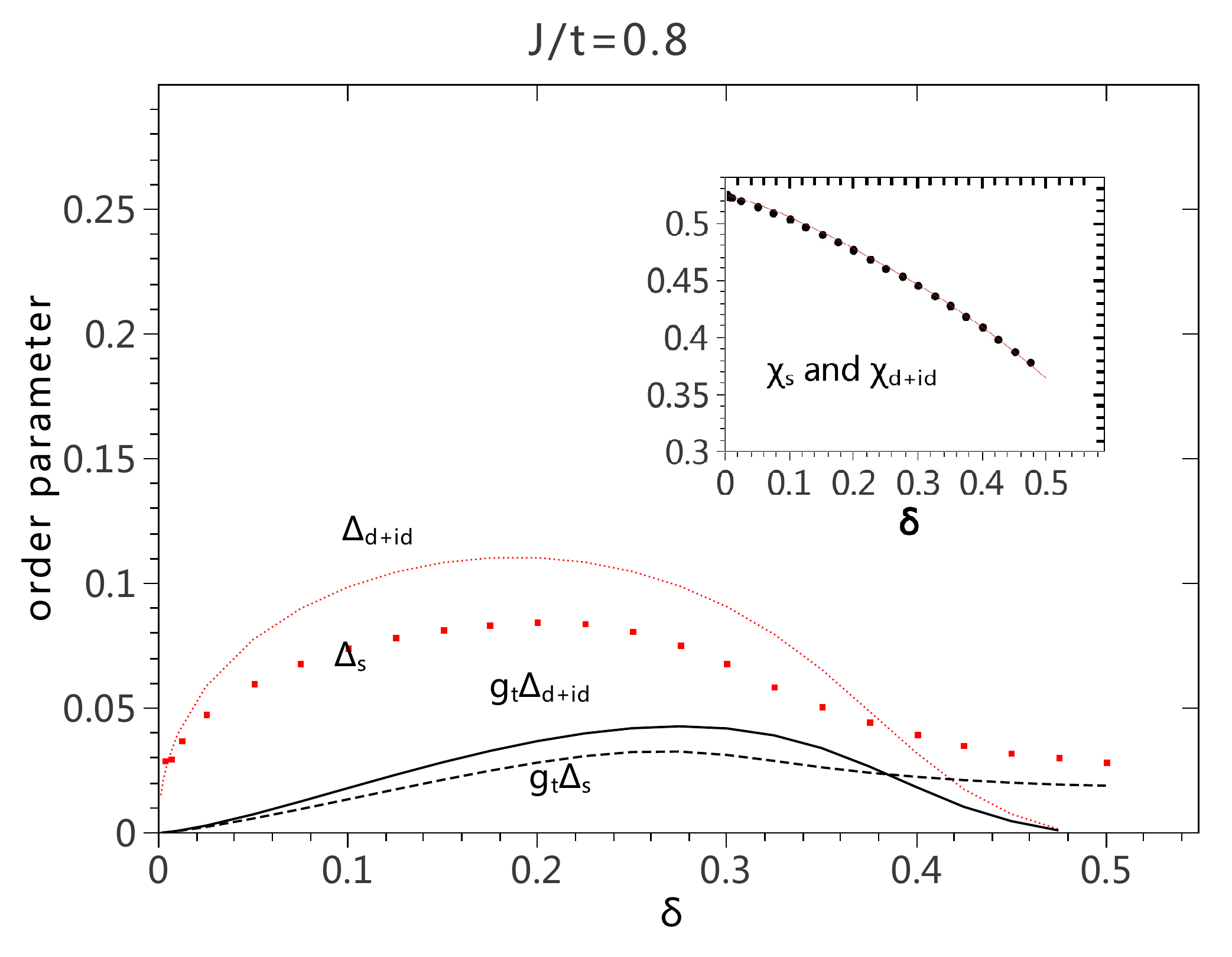}
\caption{Spin gap $\Delta_{d+id}$ and $\Delta_s$ at $T=0$ (defined in Eqs. (8) and (11)) and Superconducting Transition Temperature $(T_c\sim g_t\Delta)$ when $J_2=0$ for the $d\pm id$ and ES scenarios as a function of the hole doping parameter $\delta=1-n$; the half-filled case here corresponds to one electron per site or $\delta=0$. The order parameters are taken in units of $3g_s J/4$. Inset: one can observe that the particle-hole order parameters behave identically for the $d\pm id$ and ES situations.}
\end{figure}

\begin{eqnarray*}
E_{{\bf k}l} & = & \sqrt{|\Delta_{{\bf k}}^{i}|^{2}+|\Delta_{{\bf k}}^{I}|^{2}+\frac{1}{2}(\xi_{{\bf k}0}^{2}+\xi_{{\bf k}1}^{2})+(-1)^{l}\sqrt{\frac{1}{4}(\xi_{{\bf k}0}^{2}-\xi_{{\bf k}1}^{2})^{2}+|\Delta_{{\bf k}}^{I}|^{2}(\xi_{{\bf k}0}-\xi_{{\bf k}1})^{2}+4|\Delta_{{\bf k}}^{i}|^{2}|\Delta_{{\bf k}}^{I}|^{2}}},
\end{eqnarray*}
and for the $d\pm id$ case we use the ansatz in Eq. (12). The self-consistent equations then read,
\begin{eqnarray}
\begin{split}\frac{\partial E_{{\bf k}l}}{\partial\Delta} & =\frac{\Delta|\Gamma_{{\bf k}}^{i}|^{2}}{E_{{\bf k}l}}+\frac{\Delta|\Gamma_{{\bf k}}^{I}|^{2}}{E_{{\bf k}l}}+\frac{1}{4E_{{\bf k}l}}\frac{(-1)^{l}(16\Delta^{3}|\Gamma_{{\bf k}}^{i}|^{2}|\Gamma_{{\bf k}}^{I}|^{2}+2\Delta(\xi_{{\bf k}0}-\xi_{{\bf k}1})^{2}|\Gamma_{{\bf k}}^{I}|^{2})}{\sqrt{\frac{1}{4}(\xi_{{\bf k}0}^{2}-\xi_{{\bf k}1}^{2})^{2}+\Delta^{2}|\Gamma_{{\bf k}}^{I}|^{2}(\xi_{{\bf k}0}-\xi_{{\bf k}1})^{2}+4\Delta^{4}|\Gamma_{{\bf k}}^{i}|^{2}|\Gamma_{{\bf k}}^{I}|^{2}}}\\
\frac{\partial E_{{\bf k}l}}{\partial\mu} & =\frac{\mu}{E_{{\bf k}l}}+\frac{1}{2E_{{\bf k}l}}\frac{(-1)^{l+1}\xi_{k}(\xi_{{\bf k}0}^{2}-\xi_{{\bf k}1}^{2})}{\sqrt{\frac{1}{4}(\xi_{{\bf k}0}^{2}-\xi_{{\bf k}1}^{2})^{2}+\Delta^{2}|\Gamma_{{\bf k}}^{I}|^{2}(\xi_{{\bf k}0}-\xi_{{\bf k}1})^{2}+4\Delta^{4}|\Gamma_{k}^{i}|^{2}|\Gamma_{k}^{I}|^{2}}}\\
\frac{\partial E_{{\bf k}l}}{\partial\chi} & =\frac{-(\xi_{{\bf k}0}-\xi_{{\bf k}1})|\gamma_{{\bf k}}|}{4E_{{\bf k}l}}+\frac{0.5(-1)^{l+1}(\xi_{{\bf k}0}^{2}-\xi_{{\bf k}1}^{2})(\xi_{{\bf k}0}+\xi_{{\bf k}1})|\gamma_{{\bf k}}|-(-1)^{l}4|\Delta|^{2}|\Gamma_{{\bf k}}^{I}|^{2}|\gamma_{{\bf k}}|\xi_{{\bf k}}}{4E_{{\bf k}l}\sqrt{\frac{1}{4}(\xi_{{\bf k}0}^{2}-\xi_{{\bf k}1}^{2})^{2}+\Delta^{2}|\Gamma_{{\bf k}}^{I}|^{2}(\xi_{{\bf k}0}-\xi_{{\bf k}1})^{2}+4\Delta^{4}|\Gamma_{{\bf k}}^{i}|^{2}|\Gamma_{{\bf k}}^{I}|^{2}}}.
\end{split}
\end{eqnarray}
Note that near the superconducting transition temperature this results in 

\begin{eqnarray*}
\begin{split}\frac{\partial E_{{\bf k}l}}{\partial\Delta} & =\frac{\Delta|\Gamma_{{\bf k}}^{i}|^{2}}{E_{{\bf k}l}}+\frac{\Delta(\xi_{{\bf k}0}+\xi_{{\bf k}1})+(-1)^{l}\Delta(\xi_{{\bf k}0}-\xi_{{\bf k}1})}{E_{{\bf k}l}(\xi_{{\bf k}0}+\xi_{{\bf k}1})}|\Gamma_{{\bf k}}^{I}|^{2}=\frac{\Delta|\Gamma_{{\bf k}}^{i}|^{2}}{E_{{\bf k}l}}+\frac{\Delta}{-\mu}|\Gamma_{{\bf k}}^{I}|^{2}.
\end{split}
\end{eqnarray*}
\end{widetext}

Finally, the self-consistent mean field equations for the $d\pm id$ situation take the form 

\begin{eqnarray}
\begin{split}\delta & =-\frac{1}{N_{s}}\sum_{{\bf k}l}\tanh\frac{\beta E_{{\bf k}l}}{2}\frac{\partial E_{{\bf k}l}}{\partial\mu}\\
\Delta & =\frac{Jg_{1}}{2N_{s}}\sum_{{\bf k}l}\tanh\frac{\beta E_{{\bf k}l}}{2}\frac{\partial E_{{\bf k}l}}{\partial\Delta}\\
\chi & =\frac{J g_{1}}{2N_{s}}\sum_{{\bf k}l}\sum_{{\bf k}l}\tanh\frac{\beta E_{{\bf k}l}}{2}\frac{\partial E_{{\bf k}l}}{\partial\chi}.
\end{split}
\end{eqnarray}

Taking the limit $\Gamma_{k}^{I}=0$ is also consistent with the equations for the ES case discussed above.

\subsection{Spin Gap and Superconductivity}

\begin{figure}[ht]
\includegraphics[scale=0.7]{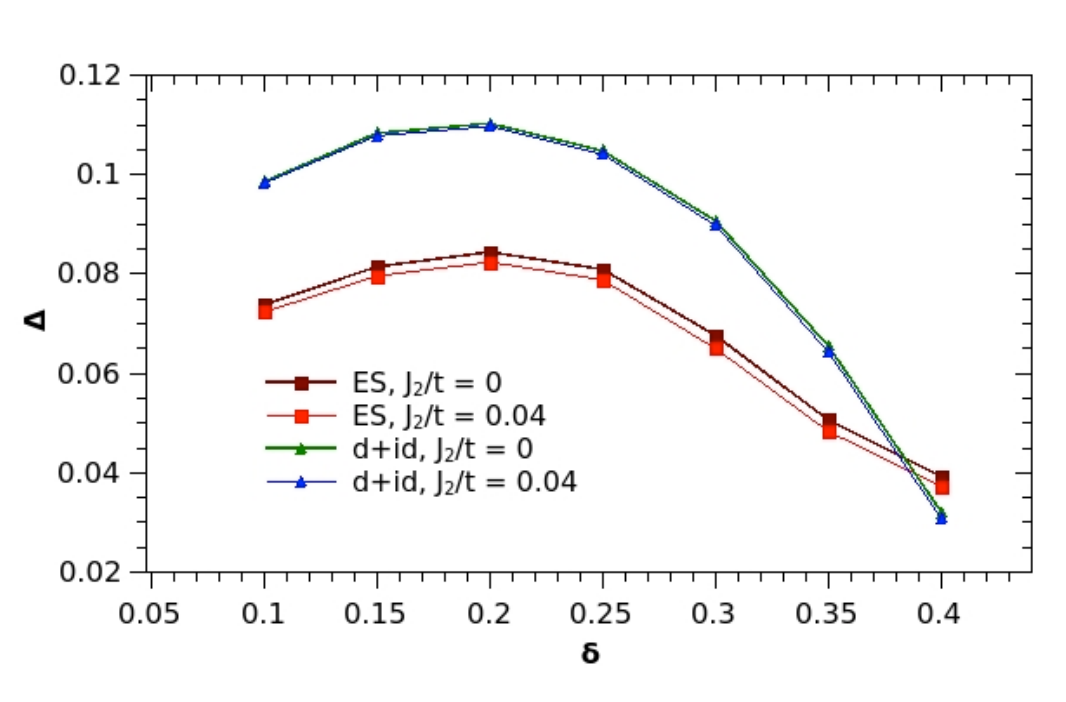}
\caption{Spin gaps (at $T=0$) versus doping, for $J_2\neq 0$, from the RMFT. Notations are similar to those in Fig. 2.}
\end{figure}

As a result of the strong on-site repulsion, the dominant pairing term is between nearest neighbor sites and counting the lattice symmetry then the natural candidate will be $d_{x^2-y^2}$ or $d_{xy}$ for nearest neighbor singlet pairing as a reminiscence of the triangular lattice \cite{Lee_p}. At a general level, one can introduce a general combination of $d_{x^2-y^2}$ and $d_{xy}$ wave pairing for the pairing term:
\begin{equation}
\label{delta}
\Delta_{{\bf k}}=\cos\theta\Delta_{x^{2}-y^{2}}({\bf k})\pm i\sin\theta\Delta_{xy}({\bf k}). 
\end{equation}

As already elaborated in Ref. \onlinecite{BlackSchaffer}, one can show that the minimum of free energy occurs for $\theta=\pi/3$. Such a pairing solution could
favor a stable ${\cal Z}_2$ spin gapped phase at half-filling for not too small $J_2$, as supported in Refs. \cite{Ran,Spin,Clark}. The ${\cal Z}_2$ symmetry can be seen from the mean-field Eq. (\ref{pairing}) following similar arguments as in Refs. \onlinecite{Ran,Spin}. It is perhaps important to underline that ${\cal Z}_2$ (gapped) spin liquids are stable in two dimensions beyond mean-field arguments \cite{Kogut} in contrast to certain $U(1)$ analogues \cite{RachelLeHur,LeeLee,Polyakov}. Other versions of spin liquids might be protected from gauge fields in the large $N$ limit \cite{IoffeLarkin,Appelquist,Hermele2,SSL2}. On the other hand, for $J_2\rightarrow 0$, the ground state at half-filling is an antiferromagnet \cite{Hohenadler,RachelLeHur,Sorella2}, and the proximity to antiferromagnetism will be addressed thoroughly through the fRG.

In Fig. 2, we present our results for the pairing strengths of the ES and $d\pm id$ situations close to half-filling when $J_2=0$. Within the RMFT, the quantity $g_t \Delta$ can be interpreted as an ``approximate'' superconducting transition temperature \cite{Gros,Anderson2,Gros2,Ogata} (remember that this approach ignores the possibility of antiferromagnetic order). As already anticipated earlier, we confirm that the $d\pm id$ solution is more favorable than the ES scenario. For completeness, we compare our results for the spin gap (RVB gap) and superconducting transition temperature at $J_2=0$ with those obtained within the U(1) slave boson approach adapted to the honeycomb lattice; consult Appendix A for a comparison with the slave-boson theory. Results obtained via the slave-boson theory are in qualitative agreement with the RMFT. By doping with holes, in the strong coupling limit, we then confirm the occurrence of a $d\pm id$ superconducting ground state; on the other hand, a more refined (probably numerical) approach would be necessary to estimate the evolution of the ground state to the heavily doped case in the case of strong interactions. Let us emphasize that for weak interactions, a $d\pm id$ superconducting ground state breaking time-reversal symmetry has been found close to 3/8 filling $(\delta=1/4)$ \cite{NCL,Kiesel}.

As shown in Fig. 3, the relative strength of the ES gap becomes less pronounced when including the effect of the finite next nearest-neighbor spin coupling $J_2$, making the superconducting transition towards the $d\pm id$ ground state more favorable. On the other hand, the RMFT ignores the presence of long-range antiferromagnetism and therefore the competition between superconductivity and antiferromagnetism will be studied via the fRG.

\section{Results from  fRG}

\begin{figure}[t!]
\centering
\includegraphics[height=.45\columnwidth]{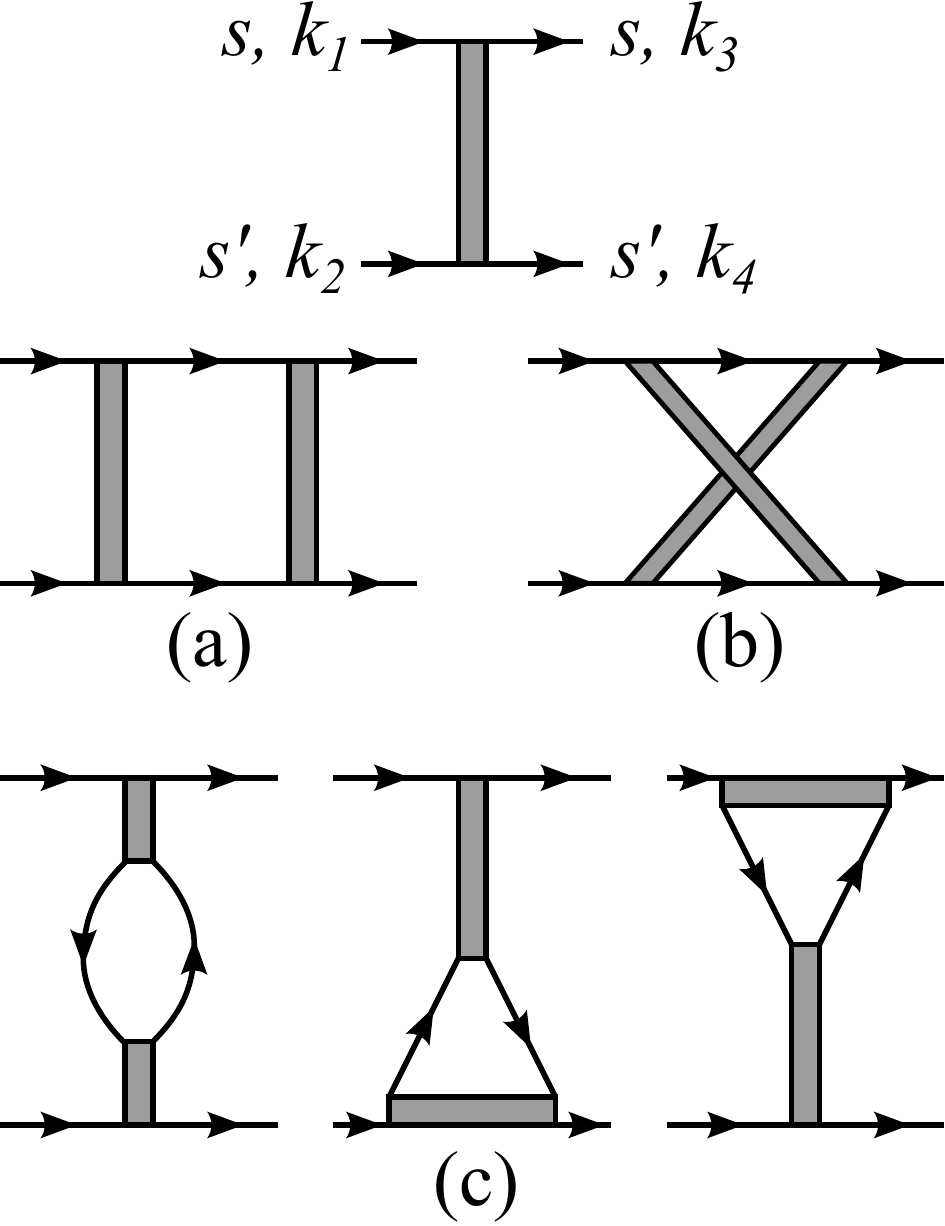} 
 \includegraphics[height=.45\columnwidth]{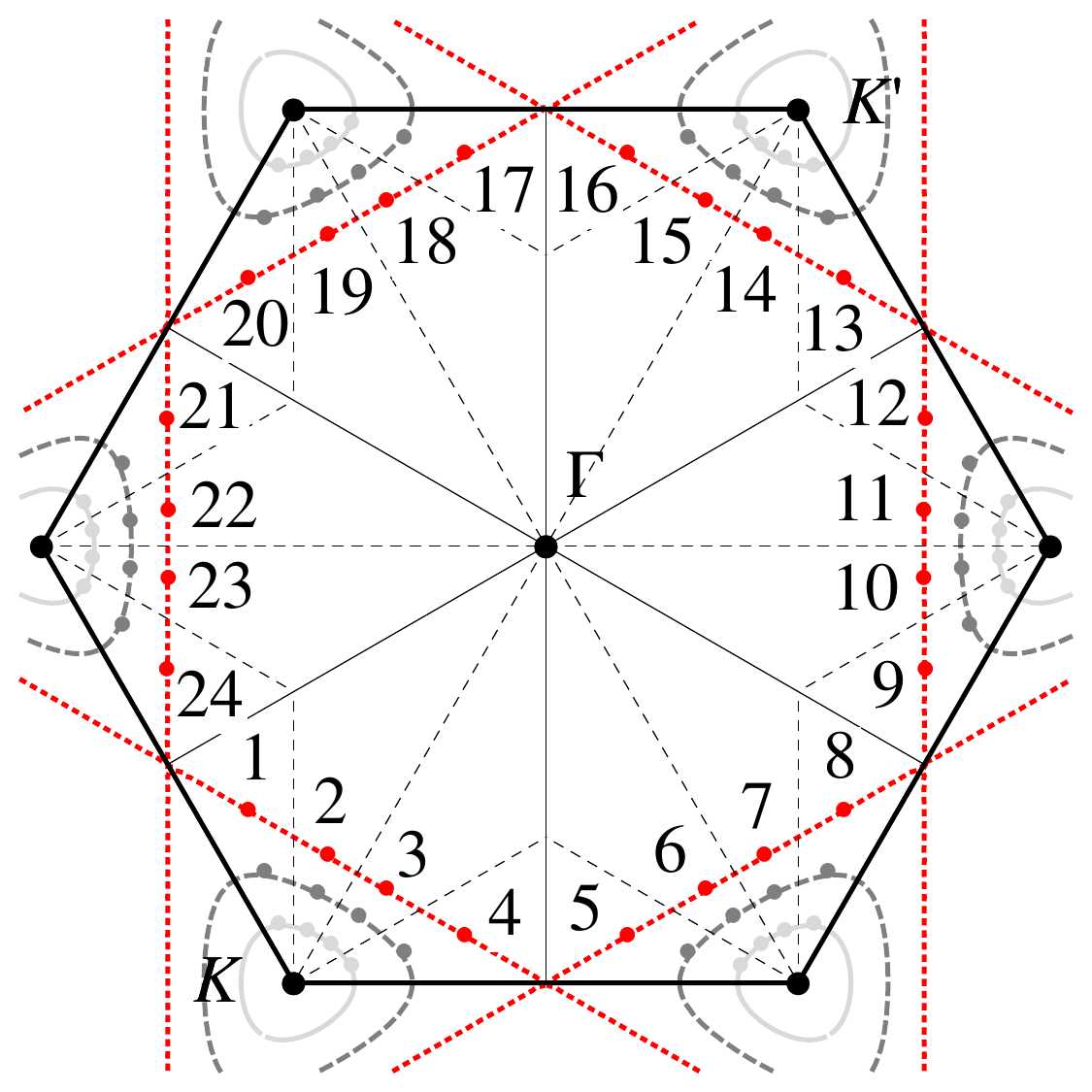}
 \caption{Left panel: Interaction vertex labeled with the spin convention (upper diagram). Below, the loop contributions to the flow of the interaction vertex including the particle-particle diagram (a), the crossed particle-hole diagram (b), and the direct particle-hole diagram (c). Right panel: ``$N=24$''-patching scheme of the Brillouin zone with constant wave vector dependence within one patch and the representative wavevector chosen on the Fermi-line shown here for three different choices of the doping $\delta$. The red dashed line corresponds to van Hove filling.}
\label{Npatch}
\end{figure}

Here, we want to describe what information can be gained beyond the RMFT by using a fRG analysis of the unconstrained $J_1$-$J_2$ model. To this end, we adapt the fRG approach for interacting fermions in the the so-called $N$-patch approximation (for a recent review, see Ref. \onlinecite{metzner2011}) to study the leading instabilities of the $J_1$-$J_2$ model on the honeycomb lattice; see Fig. \ref{Npatch}.

\subsection{Methodology}

The fRG treatment offers {\em a)} an unbiased comparison of the different possible instabilities or ordering tendencies, {\em b)} provides estimates for energy scales of these instabilities, both including the coupling of different fluctuations beyond mean-field theory.
Note however, in contrast with the previous sections, we study the unconstrained model and do not use the Gutzwiller projection. The reason for this difference is that the fRG is a technique that is perturbative in the interactions, and therefore the weakly doped situation in the Gutzwiller approach with the small renormalized hopping term $\sim \delta$ is not a good starting point for this method. So, in principle, the fRG approach for the unconstrained $J_1-J_2$ model does know about the antiferromagnetic spin interactions on neighbored sites, but not about the strong onsite correlations. In order to make up for this, we also include a moderate local Hubbard interaction and check whether our results depend qualitatively on this.

The fRG scheme employed here is the same was as recently used to explore mono-\cite{Honerkamp,raghu}, bi-\cite{scherer1,lang2012,Kiesel} and trilayer\cite{scherer2} honeycomb models with density-density interaction terms. A Brillouin zone discretization in $N$ angular patches around the $K$ and $K'$ points is employed in order to resolve the wavevector- and band-dependence dependence of the scale-dependent interactions $V_\Lambda (\vec{k}_1,n_1,\vec{k}_2,n_2,\vec{k}_3,n_3,n_4)$ (where the $\vec{k}$s denote the incoming and outgoing wave vectors, and the $n$s the band indices of the interaction). Upon integrating out the electronic degrees of freedom with lowering an infrared cutoff energy  scale $\Lambda$, the wavevector-dependent  one-loop corrections (particle-hole and particle-particle bubbles) to the bare interactions are summed up to infinite order. The standard approximation employed in this instability analysis are, just as in many previous studies (e.g. cited in Ref. \onlinecite{metzner2011}), that the electronic self energy is ignored, and that vertices of order higher than four are not taken into account. Recent work on the square lattice Hubbard model has shown that the inclusion of the self energy does not change the conclusions \cite{giering,uebelacker}.
For sufficiently strong interactions, this leads to a flow to strong coupling at a nonzero cutoff scale $\Lambda_c$ where a part of $V_\Lambda (\vec{k}_1,n_1,\vec{k}_2,n_2,\vec{k}_3,n_3,n_4)$ seems to diverge. The scale $\Lambda_c$ can be used as an estimate for the energy scale of the ordering phenomenon suggested by this flow to strong coupling. Furthermore, the wavevector- and band-dependence of the leading terms in this divergence allows one to extract the order parameters that may actually order below this instability scale.

\subsection{Results}

\begin{figure}[t!]
\centering
 \includegraphics[width=1.0\columnwidth]{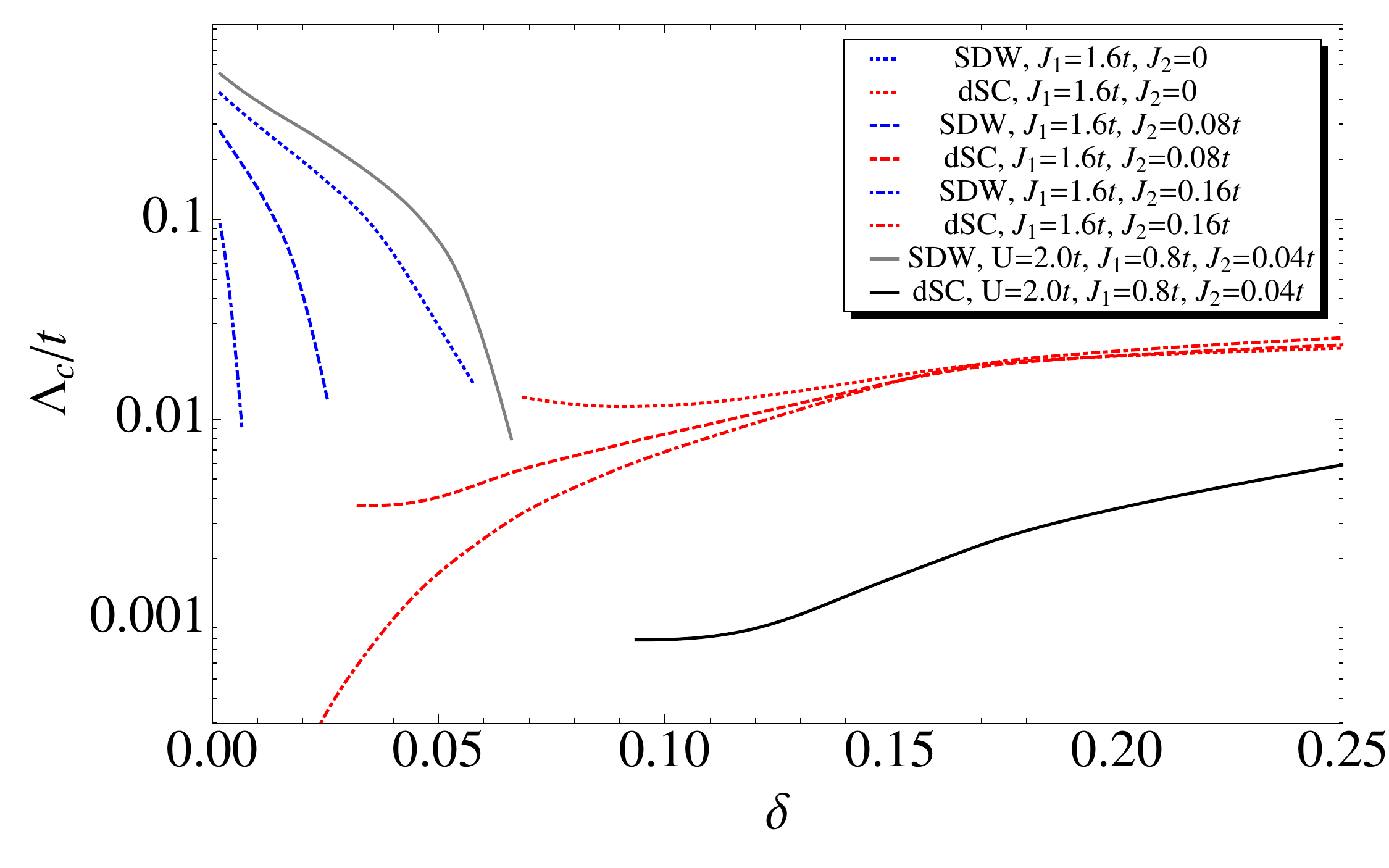}
 \caption{fRG critical scale $\Lambda_c$ in units of the hopping $t$ vs. doping $\delta$ at $T=0$ for various choices of parameters $J_1, J_2$ and onsite interaction $U$. The point $\delta =0.25$ corresponds to van Hove doping. $J_2>0$ suppresses the AF SDW ordering tendencies, but has only smaller quantitative effects on the $d$-wave pairing at larger couplings. Inclusion of $U$ does not change the qualitative findings.}
\label{J1J2delta}
\end{figure}

First let us study the case with pure spin-spin interactions and set $U=0$.
We explore a region of the phase diagram near the charge neutrality point and with $J_1 \gg J_2$. In Fig. \ref{J1J2delta} we plot the evolution of the fRG critical scale $\Lambda_c$ as function of the density deviation from half filling for $J_1=1.6t$ and three choices of small $J_2$. In the plot we also indicate the leading instability (for a description how these phases are identified from the running couplings, see, e.g., Ref. \onlinecite{Honerkamp}).
For zero and small doping, we get an antiferromagnetic (AF) spin-density-wave instability (SDW) at rather large scales. Note that due to the Dirac cones in the dispersion, a nonzero minimal interaction strength is required to obtain an instability at half  filling. The critical value comes out as $J_1 \sim 1.4t$. From the experience with the honeycomb Hubbard model with pure onsite interactions (discussed e.g. in the bilayer case in Ref. \onlinecite{lang2012}), we expect that the fRG in this approximation will somewhat underestimate the minimal true value, but qualitatively get the correct picture. The AF-SDW state was also found in QMC studies of the larger-$U$ honeycomb Hubbard model \cite{Hohenadler,CWu,Sorella2}. Hence, our study without onsite correlations fits in quite consistently.

Regarding the effect of $J_2>0$, in Fig. \ref{J1J2delta} it can be seen clearly that $J_2$ reduces the AF-SDW tendencies quite strongly. This frustration effect is of course not unexpected, and is truthfully captured by the fRG that allows for a coupling of fluctuations with different wavevectors. 

\begin{figure}[t!]
\centering 
 \includegraphics[height=.32\columnwidth]{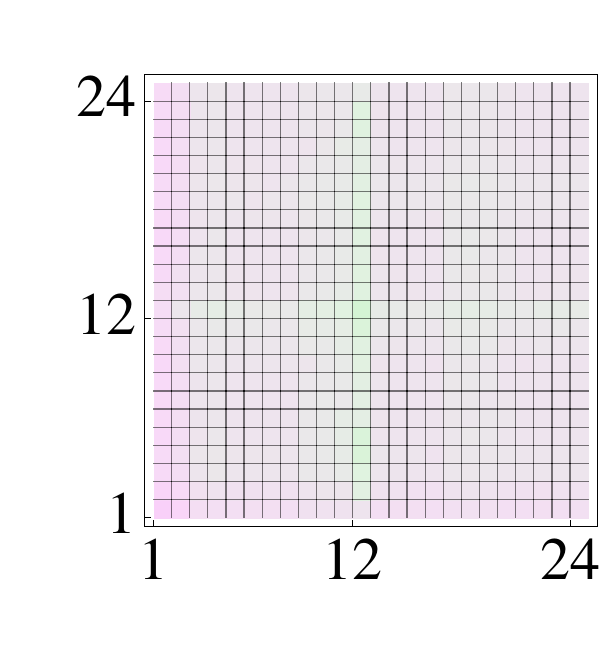}
 \includegraphics[height=.32\columnwidth]{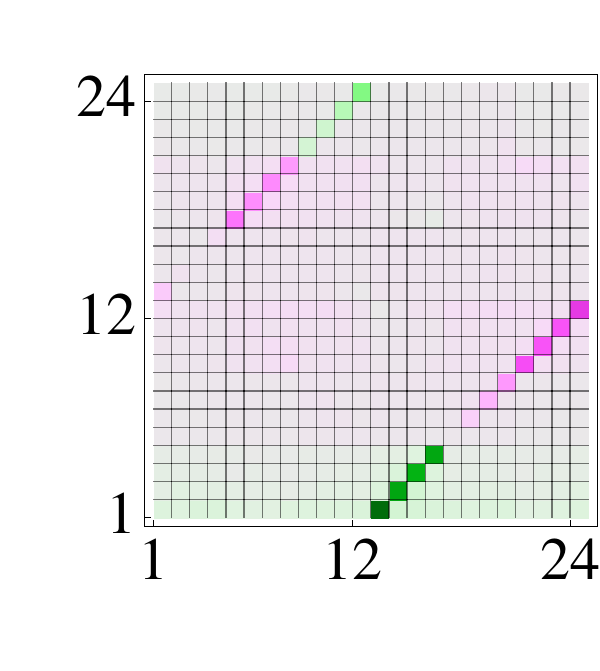}
 \includegraphics[height=.32\columnwidth]{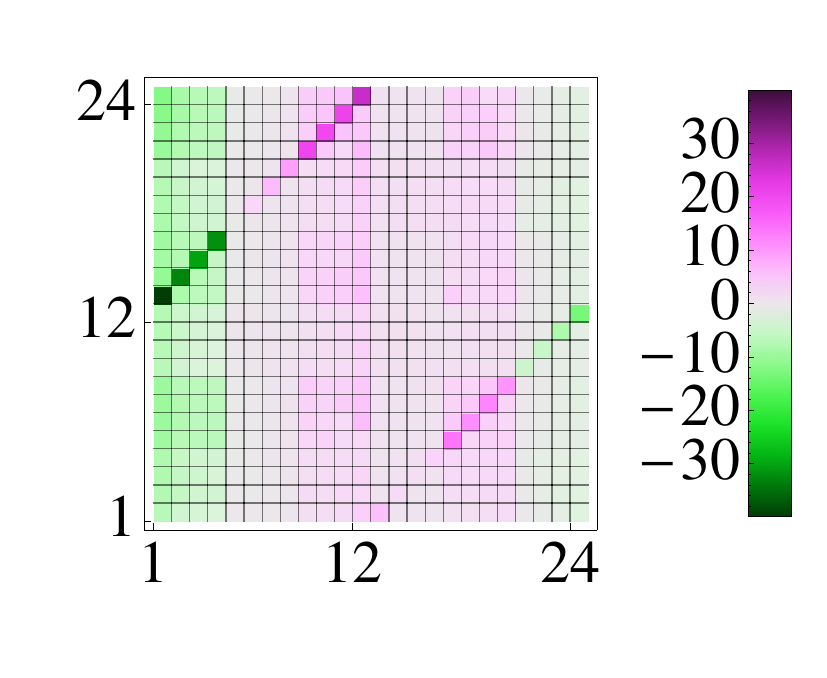}
 \caption{Typical effective interaction vertex near the critical scale in the regime of the d-wave instability in units of $t$. Left Panel: Orbital combinations with $o_1=o_2=o_3=o_4$ where $o_i \in \{a,b\}$. The numbers on the axis specify the number of the patch as shown in Fig.~\ref{Npatch}. On the horizontal axis the wavevector $k_1$ can be read off and on the vertical axis we enumerate $k_2$. $k_3$ is fixed on the first patch, $k_4$ then follows from momentum conservation. Middle Panel: Effective vertex function for the orbital combination, where $o_1=o_3, o_2=o_4\neq o_1$. Here, we can clearly identify sharp diagonal structures ($k_1=-k_2$) with a d-wave-modulation of the amplitude along the diagonal, see Fig.~\ref{pairscat}. Right panel: Effective vertex function for the orbital combination, where $o_1=o_4, o_2=o_3\neq o_1$. Also for this orbital combination a sharp diagonal structure emerges.}
\label{dwave}
\end{figure}

\begin{figure}[t!]
\centering 
 \includegraphics[height=.38\columnwidth]{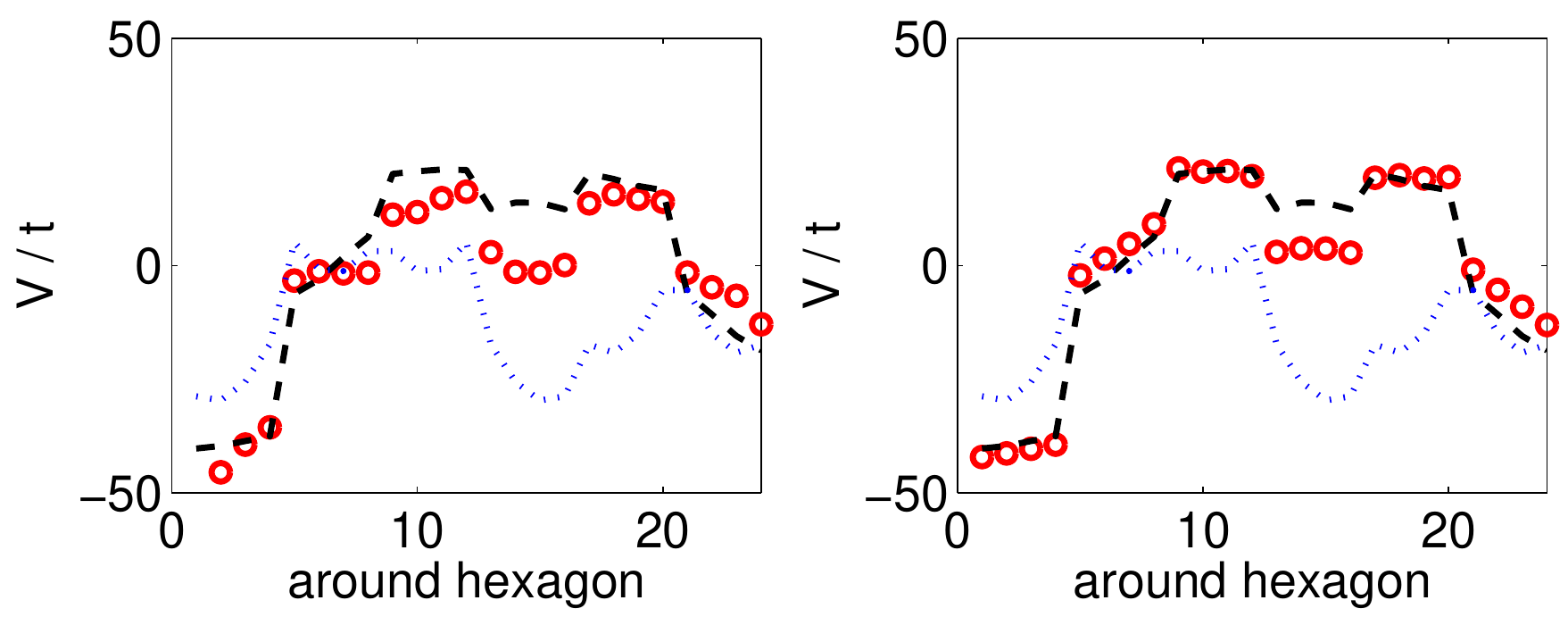}
 \caption{Pair scattering $V_\Lambda (\vec{k},-\vec{k} \to \vec{p} , -\vec{p})$ near the instability for doping $\delta=0.15$, with $\vec{k}$ fixed one one discretization point near the zone boundary, and $\vec{p}$ moving through the other points around the hexagon. The circles are the fRG data, the dashed line is the nearest-neighbor $d$-wave form factor $\propto -[ d_{x^2-y^2}^*(\vec{k}) d_{x^2-y^2}(\vec{p}) + d_{xy}^*(\vec{k}) d_{xy}(\vec{p}) ]$, and the solid line the extended $s$-wave form factor. The left plot is for $J_1=1.5t$ and $J_2=0$, the right plot for $J_1=1.5t$ and $J_2=0.01t$.}
\label{pairscat}
\end{figure}

When we increase the doping, the critical scale for the AF-SDW instability drops strongly. Then, beyond a critical doping which depends on $J_2$, a pairing instability in the $d$-wave channel takes over, at doping levels which can be inferred from  Fig. \ref{J1J2delta}. Here, for symmetry reasons, the pair scattering in the $d_{xy}$-channel and in the $d_{x^2-y^2}$-channel diverge together.  In Fig. \ref{dwave} we show snapshots of the wavevector-or patch-dependence of the effective interactions near this instability, transformed back into the sublattice basis, for different combinations of the sublattice indices. In the left panel, incoming and outgoing particles are all on the same sublattice. There, no strong interactions can be observed, i.e. the effective interaction does not have any large intra-sublattice contributions. The picture changes in the middle and right panel, where strong diagonal features with large positive and negative interactions can be found. These sharp lines occur for incoming wave vectors (labelled by the patch indices $k_1$ and $k_2$) adding up to zero, {\it i.e.}, they belong to the Cooper pair scattering channel, and the instability should be interpreted as Cooper pairing instability.  The sign structure along this line encodes the symmetry of the Cooper pair $(\vec{p},-\vec{p})$, when $\vec{p}$ moves around the Fermi surface. In order to see this symmetry more clearly, we plot in Fig. \ref{pairscat} the pair scattering $\vec{k}, -\vec{k}\to \vec{p}, -\vec{p} $ with $\vec{p}$ varying around the Fermi surface in the Brillouin zone hexagon, with $\vec{k}$ held fixed near the Brillouin zone boundary near $K$, all in the band which crosses the Fermi level. We clearly see the modulation of the pair scattering with $\vec{p}$. We also plot the $d$-wave form factor $V_{d} (\vec{k}, \vec{p}) =- V_0 \left[ d_{xy}^* (\vec{k})d_{xy}(\vec{p}) + d_{x^2-y^2}^* (\vec{k}) d_{x^2-y^2} (\vec{p})\right]$. We can see that the pair scattering in the effective interaction near the instability follows this form factor rather well, both for $J_2=0$ and for nonzero, small $J_2$.  For comparison, we also plot the form factor for extended $s$-wave pairing on nearest neighbors. This does not give any good match for the fRG data and confirms the strong dominance of the $d$-wave pairing tendencies.

In the previous sections, based on the RMFT, it was argued that the most stable pairing state in presence of the two degenerate $d$-channels would be to switch from $d+id$ at one Fermi pocket and $d-id$ at the other. The energy benefit from this comes due to the full gap now open on both Fermi circles, while a stiff $d+id$ or $d-id$ through out the BZ would have gap minima on one of the circles.  If the pair scattering between the two Fermi circles (e.g. $\vec{k}$ at $K$ and $\vec{p}$ at $K'$) is rather weak, the pair scattering will not cause a sufficient energy penalty to prevent this switching the phase of the $d$-wave superposition from one Fermi pocket to another.
With the fRG approach, without extended subsequent mean-field study of the low-energy model like e.g. in Ref. \onlinecite{platt}, it is not possible to  make any refined statements about what would be the best paired state. Note however that the pair scattering from the fRG very closely follows the simple nearest-neighbor form factor that also shows up in the mean field theory. In particular, the inter-pocket scattering between the two Fermi circles is small and of varying sign, while the scattering within one circle is stronger. In fact, in the fRG data the inter-pocket scattering is even weaker than for the nearest-neighbor form factor, which would enter the mean-field treatment. Hence, the prerequisites for switching the phase of the $d+id$ linear combination from one Fermi circle to the other to $d-id$ are all there, and the fRG supports this energy lowering. 

Note that here we do not go to larger doping where the Fermi circles get close to each other or open in the middle between the $K$ and $K'$ points. Here we expect that the near-decoupling between the $K$ and $K'$ point in the pair scattering is not valid any more, and that a unique linear combination of the degenerate $d$-basis functions needs to be chosen, as shown in Refs. \onlinecite{NCL} or \onlinecite{Kiesel}.

Let us now discuss the effect of $J_2$ on the pairing instability. As can be seen in Fig. \ref{J1J2delta}, the critical scale $\Lambda_c$ for $d$-wave pairing at small $\delta$ drops when $J_2$ is raised from 0. A natural explanation of this observation is that the $d$-wave pairing is AF-spin-fluctuation driven and hence the reduction of the AF-SDW tendencies by $J_2>0$ also reduces the pairing tendencies. However, the character of the leading instability remains unchanged by this scale change, i.e. is still of $d$ pairing type. In fact, the pair scattering near the instability follows the nearest-neighbor $d$-wave form factor even more closely than for $J_2=0$, as can also be seen in Fig. \ref{J1J2delta}.  We think that this is due to the lower scale, which implies less competition between the remnants of the SDW tendencies. Again, the extended $s$-wave pairing channel does not appear to be relevant, in agreement with the RMFT.

Furthermore, including a nonzero $U>0$ in order to account for local correlations does not change the character of the instabilities drastically. This has to be expected, as also the local repulsion drives SDW tendencies, so it basically adds to the nearest-neighbor interactions $J_1$. Correspondingly, a smaller $J_1 \sim 0.8t$ can be used to produce similar critical scales for the SDW as for $U=0$ (see Fig. \ref{J1J2delta}). On the other hand, the $d$-wave pairing instability occurs at somewhat lower scales than for $U=0$. This ties in with the earlier observation that a $U$-interaction alone with $J_1=0$ does not lead to pairing instability at reasonable scales in the slightly doped case \cite{Honerkamp}.

\section{Conclusion}

From the fRG study of the $J_1-J_2$ model on the honeycomb model, we state that the weakly coupled model exhibits a standard AF-SDW instability above a critical $J_1$ when the doping and the frustrating $J_2$ are not too large. Doping further in this regime leads to well-formed $d$-wave pairing instability, with predominant pair scattering within the respective Fermi circles  around $K$ or $K'$. Although the fRG approach dos not employ any Gutzwiller type renormalization and local correlation effects enter only perturbatively through the Hubbard $U$, these findings tie in very consistently with the RMFT approach presented in the other sections. Our results suggest that the slightly doped compound In$_3$Cu$_2$VO$_9$ \cite{Yan} could reveal an RVB phase as well as a (high-Tc) superconducting phase where the ground state is characterized by a $d+id$ singlet pairing assigned to one valley and a $d-id$ singlet pairing to the other, which then preserves time-reversal symmetry. The RMFT predicts that the $T_c$ might be quite high not too close to half-filling, as illustrated in Fig. 2.  This analysis, that takes into account the $J_2$ term, could be extended to multi-layer systems \cite{LeHurPaul,scherer1,lang2012,scherer2,Vucicevic}. We could also explore the effect of next-nearest neighbor interactions \cite{Honerkamp}.

C.H. and K.L.H. acknowledge discussions and collaborations with T. Maurice Rice.  We also acknowledge useful discussions with Doron Bergman, Silke Biermann, Andrei Chubukov, Benoit Dou\c{c}ot, Sung-Sik Lee, Tianhan Liu, Wu-Ming Liu, Stephan Rachel, Andr\' e-Marie Tremblay, Stefan Uebelacker. W. W. and K. L. H. have benefitted from the DARPA grant W911NF-10-1-0206 and from NSF grant DMR-0803200. W.W. acknowledges support from the Natural Sciences and Engineering Research Council of Canada. C.H. and M.M.S. acknowledge support from DFG FOR 723, 912 and 1162. M.M.S. is supported by the grant ERC- AdG-290623. 

\begin{appendix}

\section{Slave-Boson Theory}

Here, we provide details concerning the U(1) slave-boson theory which has been used to give Fig. 3. Extending the usual slave-particle procedure \cite{WenLee} on the honeycomb lattice, we represent the (slave) fermionic (here, spinon) operators on sublattice (A,B) respectively via $\{ c_{k\sigma}^{\dagger},c_{k\sigma}\}$, $\{ d_{k\sigma}^{\dagger},d_{k\sigma}\}$ (to build a connection with the RMFT) and the related bosonic (charge) operators $\{a_{i}^{\dagger},a_{i}\},\{b_{i}^{\dagger},b_{i}\}$. The mean-field Hamiltonian reads:
\begin{equation}
\begin{split}H=\sum_{{\bf k},\sigma} \large(\epsilon_{{\bf k}}c_{{\bf k}\sigma}^{\dagger}d_{{\bf k}\sigma}+h.c\large)-\mu_{c}\sum_{{\bf k},\sigma}\large(c_{{\bf k}\sigma}^{\dagger}c_{{\bf k}\sigma}+ d_{{\bf k}\sigma}^{\dagger}d_{{\bf k}\sigma}\large)\\
+\sum_{{\bf k}}\large(\omega_{{\bf k}}b_{{\bf k}}^{\dagger}a_{{\bf k}}+h.c\large)-\mu_{b}\sum_{{\bf k}}\large(b_{{\bf k}}^{\dagger}b_{{\bf k}}+a_{{\bf k}}^{\dagger}a_{{\bf k}}\large)\\
-\sum_{{\bf k}}\large(\Delta_{{\bf k}}(c_{{\bf k}\uparrow}^{\dagger}d_{-{\bf k}\downarrow}^{\dagger}-c_{{\bf k}\downarrow}^{\dagger}d_{-{\bf k}\uparrow}^{\dagger})+h.c.\large)\\
+N_s\large(6t\chi_{b}\chi_{c}+\frac{3J}{2}\chi_{c}^{2}+\frac{3J}{2}(1-\delta)^{2}+\frac{J}{2}\sum_{\langle ij\rangle}\Delta_{i,j}^{\dagger}\Delta_{i,j}-2\lambda\large)
\end{split}
\end{equation}
where again $N_s$ denotes the number of sites on the lattice and we have defined the order parameters (which have been chosen to be slightly different from those introduced within the RMFT):
\begin{eqnarray}
\Delta_{i,j}^{\dagger} &=& \langle c^{\dagger}_{i\uparrow}c^{\dagger}_{j\downarrow} - c^{\dagger}_{i\downarrow}c^{\dagger}_{j\uparrow}\rangle \\
\chi_c &=& \chi_d = \sum_{\sigma} \langle c^{\dagger}_{i\sigma} c_{j\sigma} \rangle \\
\chi_b &=& \chi_a = \langle b^{\dagger}_j b_i\rangle,
\end{eqnarray}
and defined $({\bf R}_{j}-{\bf R}_{i}={\bf R}_{\alpha})$
\begin{eqnarray}
\begin{split}
\epsilon_{{\bf k}} & =\Large(-t\chi_{b}-\frac{J}{2}\chi_{c}\Large)\gamma_{{\bf k}}\\
\mu_{c} & =\frac{3J}{2}(1-\delta)+\mu-\lambda\\
\omega_{{\bf k}} & =-t\chi_{c}\gamma_{{\bf k}}\\
\mu_{b} & =-\lambda\\
\Delta_{{\bf k}} & =J\sum_{\alpha}\Delta_{i,j}e^{i{\bf k}\cdot {\bf R}_{\alpha}}.
\end{split}
\end{eqnarray}
We have introduced a Lagrange multiplier $\lambda$ to reinforce the condition of half-filling, for example on sublattice A: $\lambda(a^{\dagger}_i a_i +\sum_{\sigma} c^{\dagger}_{i\sigma} c_{i\sigma} -1)$ and similarly on sublattice B. We can now proceed and diagonalize the problems by introducing a transformation for the spinons similar to Eq. (5) in the main text. An identical procedure is then applied to the bosons (chargons). 

The free energy takes the form:
\begin{equation}
\begin{split} F =-\frac{2}{\beta}\sum_{{\bf k},l}\ln(2\cosh\frac{\beta E_{{\bf k},l}}{2})-\frac{1}{\beta}\sum_{{\bf k},l}\ln(1-e^{\beta(\mu_{b}+\omega_{{\bf k}l})}) &\\
+\sum_{{\bf k},l}-\mu_{f}+(-1)^{l}\epsilon_{{\bf k}} &\\
+N_s\{6t\chi_{b}\chi_{c}+\frac{3J}{2}\chi_{c}^{2}+\frac{3J}{2}(1-\delta)^{2}+\frac{J}{2}\sum_{\langle ij\rangle}\Delta_{i,j}^{\dagger}\Delta_{i,j}-2\lambda\} &
\end{split}
\end{equation}
$l=\{0,1\}$ stems from the path integral of the two-band Hamiltonian. By analogy with the RMFT, we set
\begin{eqnarray}
\begin{split}
E_{{\bf k}l} & =\sqrt{\xi_{{\bf k}l}^{2}+|\Delta_{{\bf k}}|^{2}}\\
\xi_{{\bf k}l} & =-\mu_{c}+(-1)^{l}|\epsilon_{{\bf k}}|=-\mu_{c}+(-1)^{l}(t\chi_{b}+\frac{J}{2}\chi_{c})|\gamma_{{\bf k}}|\\
\omega_{{\bf k}l} & =(-1)^{l}|\omega_{{\bf k}}|=(-1)^{l}t\chi_{c}|\gamma_{{\bf k}}|,
\end{split}
\end{eqnarray}
giving the self-consistent equations for the ES case:
\begin{eqnarray}
\begin{split}\Delta & =\frac{\Delta}{6N_s}\sum_{{\bf k},l}\frac{J|\gamma_{{\bf k}}|^{2}}{E_{{\bf k}l}}\tanh\frac{\beta E_{{\bf k}l}}{2}\\
\chi_{c} & =\frac{1}{6N_s}\sum_{{\bf k},l}\frac{(-1)^{l}\xi_{{\bf k}l}|\gamma_{{\bf k}}|}{E_{{\bf k}l}}\tanh\frac{\beta E_{{\bf k}l}}{2}\\
\chi_{b} & =-\frac{1}{6N_s}\sum_{{\bf k},l}\frac{(-1)^{l}|\gamma_{{\bf k}}|}{e^{\beta(\omega_{{\bf k}l}+\mu_{b})}-1}.
\end{split}
\end{eqnarray}

\begin{figure}[ht]
\includegraphics[scale=0.55]{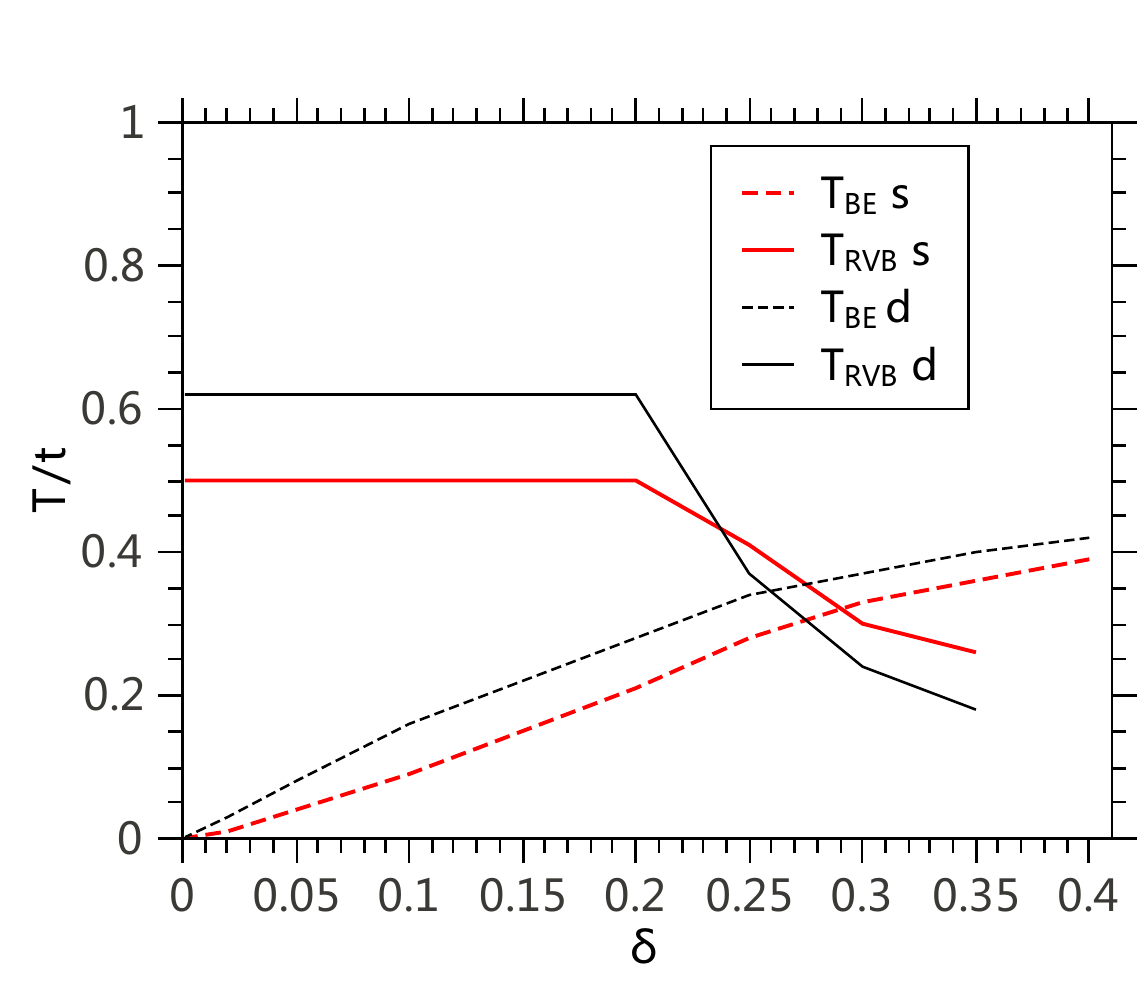}
\caption{Phase diagram obtained with the $U(1)$ slave-boson approach for the case $J_2=0$ $(here, J/t=1)$. Details of the theory are presented in Appendix A as well as the definitions of the temperature scales $T_{BE}$ and $T_{RVB}$ which correspond respectively to the temperatures associated with the boson (chargon) condensation and spin gap formation, respectively. Within this approach, an upper bound on the Superconducting Transition Temperature is given by Min$(T_{BE},T_{RVB})$.}
\end{figure}

The chemical potential for fermions $\mu_{f}$ and bosons $\lambda$ are decided
by $\sum_{\sigma}\langle  f^{\dagger}_{i\sigma} f_{i\sigma}\rangle = 1-\delta = -\frac{1}{2N_s}\frac{\partial F}{\partial\mu}$ and $\frac{\partial F}{\partial\lambda}=0$
(actually $n_{b}$ is also determined by $\mu$, implicitly. The Lagrange multiplier $\lambda$ connects the $\delta=\langle b^{\dagger}_i b_i\rangle$ and $\mu$), 
\begin{equation}
\delta=\frac{1}{2N_s}\sum_{{\bf k},l}\frac{\xi_{{\bf k}l}}{E_{{\bf k}l}}\tanh\frac{\beta E_{{\bf k}l}}{2}
\end{equation}
 and 
\begin{equation}
\delta=\frac{1}{2N_s}\sum_{{\bf k},l}\frac{1}{e^{\beta(\omega_{{\bf k}l}+\mu_{b})}-1}.
\end{equation}
To evaluate the superconducting order parameter $\langle c_{i\uparrow}^{\dagger}c_{j\downarrow}^{\dagger}-c_{i\downarrow}^{\dagger}c_{j\uparrow}^{\dagger}\rangle=\langle b_{i}b_{j}\rangle\langle f_{i\uparrow}^{\dagger}f_{j\downarrow}^{\dagger}-f_{i\downarrow}^{\dagger}f_{j\uparrow}^{\dagger}\rangle$
we can simply assume $\langle b_{i}b_{j}\rangle$ $\approx$ $\langle b_{i}\rangle\langle b_{j}\rangle\neq0$,
and $\langle f_{i\uparrow}^{\dagger}f_{j\downarrow}^{\dagger}-f_{i\downarrow}^{\dagger}f_{j\uparrow}^{\dagger}\rangle$
$\neq$ 0, $\textit{i. e.}$, we can numerical solve the self-consistent equations  to find the temperature $T_{BE}$ for holon condensation and $T_{RVB}$ for spin gap formation; see Fig. 8. 

We have checked that the self-consistent equations in Eqs. (A8) are consistent with those obtained within the RMFT in Eqs. (21) in the main text. Furthermore, we straightforwardly obtain similar equations as Eqs. (22) (in the main text) for the $d\pm id$ situation. 

\end{appendix}

\end{document}